\documentclass{article}

\usepackage{PRIMEarxiv}
\usepackage{cite}
\usepackage{amsmath,amssymb,amsfonts}
\usepackage{algorithmic}
\usepackage{graphicx}
\usepackage{siunitx}
\usepackage{textcomp}
\usepackage{upgreek}
\usepackage[utf8]{inputenc} 
\usepackage[T1]{fontenc}    
\usepackage{hyperref}       
\usepackage{url}            
\usepackage{booktabs}       
\usepackage{amsfonts}       
\usepackage{nicefrac}       
\usepackage{microtype}      
\usepackage{lipsum}
\usepackage{fancyhdr}       
\usepackage{graphicx}       
\graphicspath{{media/}}     

\title{Fast implicit diffusive dark-field retrieval for single-exposure, single-mask x-ray imaging}

\author{
 Mario A. Beltran\textsuperscript{1}, David M. Paganin\textsuperscript{1}, Michelle K. Croughan\textsuperscript{1}, Kaye S. Morgan\textsuperscript{1} \\
 \\
\textsuperscript{1}
School of Physics and
Astronomy, Monash University, Clayton, VIC 3800, Australia\\
\texttt{email: mario.beltran@monash.edu}}

\begin{document}
\maketitle

\begin{abstract}
Complementary to conventional and phase X-ray radiography, dark-field imaging has become central in visualizing diffusive scattering signal due to the spatially-unresolved texture within an object. To date most diffusive-dark-field retrieval methods require either the acquisition of multiple images at the cost of higher radiation dose or significant amounts of computational memory and time. In this work, a simple method of X-ray diffusive dark-field retrieval is presented, applicable to any single-mask imaging setup, with only one exposure of the sample. The approach, which is based on a model of geometric and diffusive reverse-flow conservation, is implicit and non-iterative. This numerically fast methodology is applied to experimental X-ray images acquired using both a random mask and a grid mask, giving high quality reconstructions that are very stable in the presence of noise.  The method should be useful for high-speed imaging and/or imaging with low-flux sources.
\end{abstract}


\section{Introduction}\label{sec1}

Attenuation-based X-ray radiography is a powerful non-invasive technique to study samples at sub-millimeter length scales. However, when the difference in absorption of the transmitted beam by adjacent materials is small, the sample features become difficult to visualize. Hence, absorption-based X-ray radiography has limited applications, encountering difficulties when imaging weakly absorbing specimens without the use of contrast enhancement agents \cite{Bravin2013}.

In recent years, complementary phase (refraction) and diffusive dark-field signals have also been acquired using X-ray phase-contrast imaging set-ups. These include analyzer-based systems \cite{FosterABI1980}, grating interferometers \cite{momose2003,WeitKamp2005,FranzPfeiffer2006LowBrilliancePC,PFEIFERgrating:2006}, edge-illumination set-ups \cite{OLIVOedge:2007,Endrizzi2014}, single-mask imaging (grating or random) \cite{Wen2010,KMorganGrid:2011,BerujonSpeckle:2012, KMorgan:2012} and propagation-based imaging (PBI) \cite{Cloetens1996,SnigerevPBI1995,SWILKINS1996}. Here we use a single-mask imaging set-up, where the diffusive dark-field signal has been shown to be successfully extracted \cite{Berujon2010Multi,Morgan2013Grid,Zanette:2014,Zdora:UMPA}. Such a dark-field signal often stems from small-angle or ultra-small-angle X-ray scattering (SAXS/USAXS) from unresolved texture within the object, revealing features that would otherwise not be revealed in either attenuation or phase contrast X-ray images. Therefore diffusive dark-field signal extraction has potential use in multiple fields, such as biomedical studies, material science and industrial inspection. Further to this, in recent years dark-field retrieval methods have been extended to recover directional diffusive dark-field signal, enabling the orientation of SAXS/USAXS-producing texture or fibers to also be visualized \cite{TorbenDDF2010,Jensen2010b,TUNHE2018,FlorianSchaff2017, dreier2020, smith2022}. In the single-mask approach, a mask patterns the X-ray intensity incident on the sample, then an image captured some distance downstream of the sample reveals how it has distorted this pattern. Local transverse shifts of the pattern reveal phase gradients introduced by the sample and a local blurring of the pattern reveals micro-structure in the sample.     

Methods used to extract the diffusive dark-field signal from single-mask set-ups usually require some kind of pixel-by-pixel analysis, which can lead to large computation times. To improve spatial resolution, some approaches acquire exposures at multiple translational mask positions \cite{berujon2015, Zdora:UMPA}, at the cost of higher radiation dose. Diffusive dark-field retrieval methods can be classified into two categories, explicit or implicit. Explicit methods directly track the distortions of the mask features, via a suitable form of correlation analysis or iterative computational approach, to retrieve the phase and diffusive dark-field signals \cite{Berujon2010Multi,YingAndMorgan:2022,Zdora:UMPA,Zanette:2014}. Implicit methods indirectly track mask distortions, extracting the signals in a deterministic manner via direct solution of a suitable partial differential equation. This was proposed by Paganin $et$ $al$.~\cite{Paganin:2018} for phase imaging, founded on the concept of local optical energy conservation and geometric flow to extract phase information, assuming the sample to be fully transparent. This algorithm was later extended by Pavlov $et$ $al$.~\cite{Pavlov:2020} and Qu\'{e}not $et$ $al$.~\cite{LaureneQuenot:2021} to include sample attenuation effects. The idea was taken further by Pavlov $et$ $al$.~\cite{Pavlov:2020B,PavlovDdark:2021} and Alloo $et$ $al$.~\cite{SAlloo:2022}, to recover a diffusive dark-field signal. As a starting point, their approach utilized the Fokker-Planck generalization \cite{PaganinMorgan:2019,MorganPaganin:2019} of the transport-of-intensity equation of paraxial wave optics \cite{Teague1983}, adapted to random mask-based imaging. The dark-field retrieval algorithm of Pavlov {\em et al.}~and Alloo {\em et al.}~requires a minimum of two sample and two reference images, captured at different transverse positions of the mask. 

In this paper we present an implicit diffusive-dark-field reverse-flow retrieval method that is also inspired by the Fokker-Planck equation.  Our method requires only a single exposure of the mask and a single exposure of the mask-and-sample, greatly reducing the complexity and duration of the acquisition process and drastically improving computation time. The theoretical forward model and procedure is presented in Section~\ref{sec2}, followed by an experimental verification in Section~\ref{sec3}. Section~\ref{sec4} discusses our key findings, followed by some concluding remarks. 

\section{Theory} \label{sec2}

\begin{figure}[h!]
\centering\includegraphics[width=13cm]{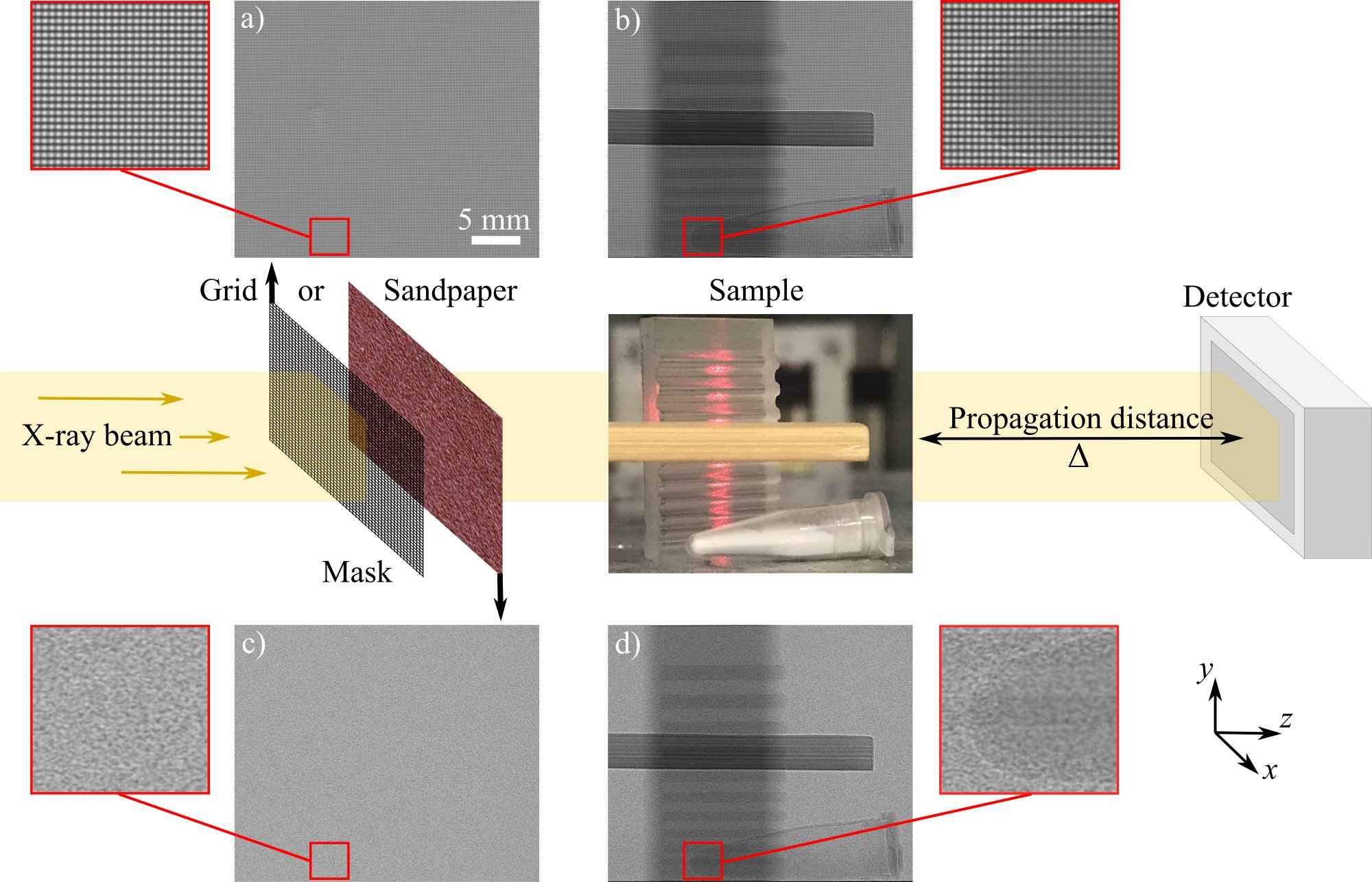}
\caption{Schematic illustrating our single-mask imaging setup. Exposures of a periodic grid mask are shown in (a) and (b), without and with a sample present. Exposures of a piece of sandpaper are shown in (c) and (d), without and with a sample present. The scale bar in (a) also applies to the images in (b-d). The sample here consists of a shaped PMMA block, a PMMA tube containing polystyrene microspheres and a rectangular wooden skewer.}\label{Fig1}
\end{figure}

Consider Fig.~\ref{Fig1}, where paraxial forward-propagating X-rays illuminate a mask, before passing through a sample and then traversing a distance $\Delta$ to the planar surface of a position-sensitive detector. Let $\textup{I}_{R}$ be the recorded intensity of the reference mask, i.e., the image taken in the absence of the sample, as a function of Cartesian coordinates $\textbf{r}= \left ( x , y \right )$ in a plane perpendicular to the optical axis $z$. The recorded image in the presence of both the sample and mask is denoted by $\textup{I}_{S}$. Assume the sample to obey the projection approximation \cite{Paganin2006} at sufficiently high X-ray energies.  The presence of the sample will distort the illumination pattern created by the reference X-ray mask. These distortions are physically manifested as: (i) Geometric flow, where certain features of the reference image are transversely displaced due to the refraction of rays induced by the sample, in a manner that locally conserves the total radiant exposure; (ii) Flux loss, where the brightness of the reference image is diminished as a result of the attenuation of rays as they traverse the sample; (iii) Flux-conservative diffusion, where some features of the reference pattern are blurred out as a consequence of SAXS/USAXS or unresolved multiple-scattering of the beam, often occurring when the rays interact with micro-structures within the volume.

We have the freedom to choose whether we consider $\textup{I}_{R}$ as evolving into $\textup{I}_{S}$ due to the introduction of the sample, or the reverse-flow case where $\textup{I}_{S}$ devolves into $\textup{I}_{R}$ due to the removal of the sample.  Physically, these two viewpoints are equivalent, but the logical possibility exists that they are inequivalent from the perspective of the inverse problem of retrieving flow, flux loss and diffusion. As we now show, the reverse-flow perspective enables us to solve the inverse problem in a surprisingly simple manner.  

Treating the devolution-type distortions from $\textup{I}_{S}$ to $\textup{I}_{R}$ in a reverse-evolution Fokker-Planck-type manner \cite{PaganinMorgan:2019}, we write down a continuity-type equation incorporating terms that describe geometric flow, flux loss and diffusion (cf.~Refs.~\cite{Paganin:2018,PavlovDdark:2021,Pavlov:2020,Pavlov:2020B}):  
\begin{equation}
\overset{{\textup{Flux}}}{\overbrace{\textup{I}_{S}  -  \textup{I}_{R}\textup{t}_{0}}}  =  \overset{{\textup{Diffusion}}}{\overbrace{\Delta ^{2}\nabla_{\perp}^{2}\left (\mathfrak{D}\textup{I}_{S}  \right )}}  -  \overset{{\textup{Flow}}}{\overbrace{\nabla_{\perp }\cdot \left (  \textup{I}_{S}\textbf{S}_{0} \right )}}.\label{eq:EQ01}
\end{equation}
Here, $\textup{I}_{S}$, $\textup{I}_{R}$, $\textup{t}_{0}$, $\mathfrak{D}$ and $\textbf{S}_{0}$ are all dependent on the transverse coordinates $\textbf{r}$, with the transverse gradient and Laplacian being denoted as $\nabla_{\perp}= (\partial_{x},\partial_{y})$ and $\nabla_{\perp}^{2}= \partial_{x}^{2} + \partial_{y}^{2}$, respectively. The dimensionless transmission function $\textup{t}_{0}$ quantifies the attenuation properties of the object, yielding a flux loss in the reference pattern \cite{PavlovDdark:2021,SAlloo:2022}. $\mathfrak{D}$ is the dimensionless diffusive function \cite{PaganinPelliccia2020,Paganin2022Preprint}
\begin{equation}
\label{eq:PositionDependentScalarDiffusionCoefficient}
    \mathfrak{D} = \tfrac{1}{2}\theta_S^2F
\end{equation}
due to SAXS/USAXS which we wish to recover \cite{PaganinMorgan:2019,MorganPaganin:2019}, with $\theta_S$ being the position-dependent blur-cone half-angle of the local SAXS/USAXS diffuse scatter that arises due to the interaction of the X-rays with the spatially-unresolved random microstructure in the sample, and $F$ being the position-dependent proportion of the incident rays that are converted to diffuse scatter. The vector function $\textbf{S}_{0} = (\textup{S}_{0x},\textup{S}_{0y})$, denotes the transverse displacements (geometric flow \cite{Paganin:2018}) of the reference pattern upon introduction of the sample, where each component has units of distance. Under coherent illumination, $\textbf{S}_{0}$ is related to gradients in the X-ray wavefield phase $\phi_0$ introduced by the sample, via \cite{KMorganGrid:2011}
\begin{equation}
\textbf{S}_{0} = \frac{\Delta  }{k}\nabla_{\perp }\phi_{0},\label{eq:EQPhase}
\end{equation} 
\noindent where the wavenumber $k = 2\pi/\lambda$ comes from the wavelength $\lambda$ of the radiation. Note, also, that no particular functional form is assumed for the previously-mentioned position-dependent blur cone associated with the diffuse scatter from each transverse location in the sample \cite{Paganin2022Preprint}.

From this point, we restrict ourselves to objects comprised of a single material of possibly varying density, that is, materials whose position-dependent complex refractive index $n= 1-\delta+i\beta$ is such that the ratio $\delta / \beta$ is everywhere the same, within the volume occupied by the sample \cite{Paganin:2002,Paganin:2004}. Under this assumption, the object transmission $\textup{t}_{0}$ can be recovered using the method of Pavlov $et$ $al$.~\cite{Pavlov:2020}, which is based on the transport-of-intensity equation \cite{Teague1983} (cf.~the method of Ref.~\cite{Paganin:2002}):
\begin{equation}
\textup{t}_{0} \approx \mathcal{F}^{-1}\left (  \frac{1}{1+\frac{\Delta \delta}{\mu} \left | \textbf{q} \right |^{2}} \mathcal{F}\left \{ \frac{\textup{I}_{S}}{\textup{I}_{R}} \right \}  \right ).\label{eq:EQ02}
\end{equation}  
\noindent Here, $\mu= 2k\beta$ is the object's linear attenuation coefficient, $\mathcal{F}$ denotes Fourier transformation with respect to $\textbf{r}$, $\mathcal{F}^{-1}$ denotes the corresponding inverse transformation, $\textbf{q}= \left ( q_{x},q_{y} \right )$ are transverse Fourier space coordinates reciprocal to $\textbf{r}$, and the utilized Fourier-transform convention and notation is 
%
\begin{subequations}
\begin{equation}
 \mathcal{F}[g\left ( \textbf{r} \right )] \equiv \widetilde{g}\left ( \textbf{q} \right )= \frac{1}{2\pi}\iint_{\infty }^{\infty }e^{-i\textbf{q}\cdot \textbf{r}}g\left ( \textbf{r} \right )d\textbf{r}
 \label{eq:EQ03A}, \\  
\end{equation}
\begin{equation}
\!\!\!\!\!\!\!\!\mathcal{F}^{-1}[\widetilde{g}\left ( \textbf{q} \right )] \equiv g\left ( \textbf{r} \right ) = \frac{1}{2\pi}\iint_{\infty }^{\infty }e^{i\textbf{q}\cdot \textbf{r}}  \widetilde{g}\left ( \textbf{q} \right )d\textbf{q} \label{eq:EQ03B}.
\end{equation}
\end{subequations}

Although the single-material assumption seems highly restrictive, it has been demonstrated in numerous practical X-ray phase contrast studies that retrieval methods work effectively for samples that stretch beyond this assumption, even to samples containing multiple materials (see e.g.~\cite{Gureyev2materials:2002,Beltran:2010, irvine2014}). For high-energy X-rays and low-atomic-number materials, there is a simple physical reason why different elements may be considered as a single material for PBI purposes.  This single material is electrons, because the influence of atomic nuclei may often be ignored in such a regime, with electron density dominating the photon--matter interaction. 

\begin{figure}[h!]
\centering\includegraphics[width=13cm]{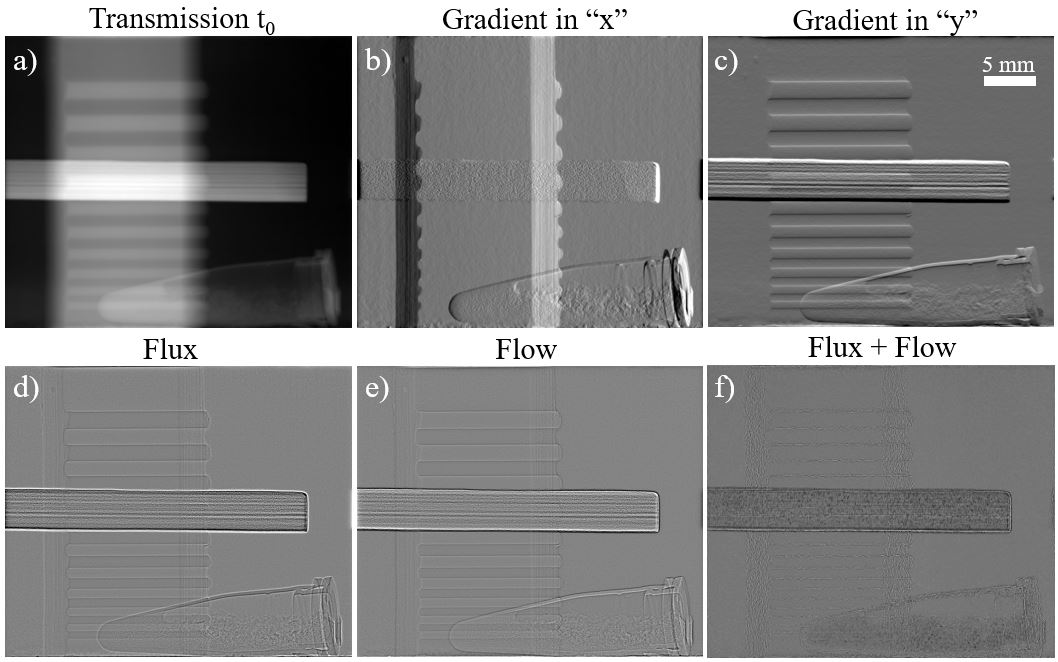}
\caption{Processed images used as inputs to calculate the ``Flux'', ``Flow'' and ``Diffusion'', with the particular sample discussed in Section~\ref{sec3}. (a) Logarithm of transmission, $-\textup{ln}[\textup{t}_{0}]$, recovered using Eq.~(\ref{eq:EQ02}). (b) Derivative in the $x$-direction, $\partial_{x} \textup{ln}[\textup{t}_{0}]$, calculated from (a). (c) Derivative in the $y$-direction, $\partial _{y} \textup{ln}[\textup{t}_{0}]$, also calculated from (a). (d) Image of calculated ``Flux''. (e) Image of calculated ``Flow''. (f) ``Flux+Flow'' image. The scale bar in (c) applies to all panels.}\label{Fig2}
\end{figure}

In Fig.~\ref{Fig2}(a) we show the recovered transmission, $\textup{t}_{0}$, found by applying Eq.~(\ref{eq:EQ02}) to the $\textup{I}_R$ and $\textup{I}_S$ images shown in Figs.~\ref{Fig1}(c) and \ref{Fig1}(d) respectively. The experimental details describing how the images were captured are discussed in Section~\ref{sec3}, and we concentrate here on describing the retrieval procedure. Once the sample transmission, $\textup{t}_{0}$, has been obtained, we are now able to easily calculate the ``Flux'' term in Eq.~(\ref{eq:EQ01}), with the resulting image shown in Fig.~\ref{Fig2}(d).

Turning attention to the ``Flow'' term, in the case of a single-material sample this can be re-expressed in terms of the transmission using the phase--intensity relation $\phi_{0} = \frac{k\delta}{\mu}\textup{ln}(\textup{t}_{0})$ \cite{Paganin:2002}, hence
\begin{eqnarray} 
\overset{\textup{Flow}}{\overbrace{\frac{\Delta  }{k}\nabla_{\perp }\cdot \left ( \textup{I}_{S}\nabla_{\perp }\phi_{0} \right )}}&=&  \frac{\Delta \delta }{\mu}\nabla_{\perp }\cdot \left (\textup{I}_{S}\nabla_{\perp }\textup{ln}[\textup{t}_{0}] \right )  \nonumber\\
&=& \frac{\Delta \delta }{\mu}\left \{\partial _{x}\left ( \textup{I}_{S}\partial _{x} \textup{ln}[\textup{t}_{0}]
\right )+\partial _{y}\left ( \textup{I}_{S}\partial _{y} \textup{ln}[\textup{t}_{0}]
\right )  \right \}.\label{eq:EQ04}
\end{eqnarray}
\noindent  The Fourier operators $\partial _{x}= \mathcal{F}^{-1}iq_{x}\mathcal{F}$ and $\partial _{y}= \mathcal{F}^{-1}iq_{y}\mathcal{F}$ \cite{Paganin2006} can be used to compute the transverse spatial derivatives. This gives the calculated derivatives in the $x$- and $y$-directions of the transmission in Fig.~\ref{Fig2}~(a), as shown in Figs.~\ref{Fig2}(b) and \ref{Fig2}(c), respectively. These are then subsequently used to compute the ``Flow'' term seen in Fig.~\ref{Fig2}(e).   

Having obtained the ``Flux'' and ``Flow'' we can simply add these to get a ``Flux+Flow'' image (see Fig.~\ref{Fig2}(f)), and then apply an inverse Laplacian operation to solve for the diffusive dark-field $\mathfrak{D}$. Hence
\begin{equation}
\mathfrak{D} =\frac{\nabla_{\perp}^{-2} \left [ { \textup{Flux}}+{\textup{Flow}} \right ]}{\Delta ^{2}\textup{I}_{S}}.\label{eq:EQ05}
\end{equation}
The inverse Laplacian can be efficiently numerically implemented using 
\begin{equation}
\nabla^{-2}= -\mathcal{F}^{-1}\frac{1}{\left | \textbf{q} \right |^{2}+\varepsilon }\mathcal{F}.\label{eq:EQ06}
\end{equation}
The added Tikhonov regularizing term $\varepsilon>0$ is small compared to $\left | \textbf{q} \right |^{2}$, with the exception of the origin of Fourier space, in order to avoid a division-by-zero singularity when $(q_{x},q_{y})= (0,0)$. 

\section{Methods and results}\label{sec3}

To test the approach described in Section~\ref{sec2}, X-ray images were collected in hutch 3B of the Imaging and Medical Beamline (IMBL) at the Australian Synchrotron. The setup of our experiment is the same as a typical single-mask imaging technique, as illustrated in Fig.~\ref{Fig1}. A large source-to-object distance (135 m) and  double-crystal monochromator provided a near planar X-ray beam. We used 25 keV X-rays to produce visible phase contrast fringes and diffusive dark-field effects for these small samples. The beam size was collimated to approximately 31.5 mm wide and 26.5 mm high, which was large enough to illuminate each sample. A 25 $\upmu$m Gadox scintillator and lens-coupled PCO.Edge camera, producing an effective pixel size of 12.3 $\upmu$m, was used to collect the images.

The masks were positioned 0.5 m upstream from the object (as close as possible) and the detector at a distance $\Delta= 2$ m downstream from the object. This relatively intermediate sample-to-detector propagation distance was sufficient to render visible the blurring of the reference pattern from the diffusive dark-field-producing parts of the sample. Each image was acquired with an exposure time of 1 s. Flat field images (with no object in the beam) and dark current images (with no incident X-rays) were recorded to normalize the image intensity, correct for the beam non-uniformities and correct for pixel-to-pixel variations in detector sensitivity. 

For the periodic-mask experiments, an attenuating grid was used, in this case a geological stainless steel sieve with holes of 90 $\upmu$m and wires 61 $\upmu$m thick, which attenuates 0-72\% of the 25 keV X-rays. In the case of a random mask, 3 sheets of sandpaper were used as an analyzer mask. 

The first sample was (i) a plastic tube filled with polystyrene microspheres of around 6 $\upmu$m in diameter and (ii) a wooden skewer, placed in front of (iii) a smoothly undulating PMMA plastic slab of approximately 7 mm mean thickness (see center of Fig.~\ref{Fig1}). For a second sample, a slice of lemon was mounted with some wooden craft sticks (seen in the top-right of the images).  

\subsection{Diffusive dark-field imaging using a single random mask}\label{subsec1}

In this section we demonstrate diffusive dark-field recovery using the methodology described in Section~\ref{sec2}.  This is applied to images acquired using a random mask (sandpaper) as a reference mask, namely the raw data displayed in Figs.~\ref{Fig1}(c) and \ref{Fig1}(d). For a beam of 25 keV X-rays, effective values of $\delta =  4.265\times 10 ^{7}$ and $\beta = 1.53\times 10 ^{10}$, specific for water, were used \cite{Bravin2013}. This is consistent with the $\delta/\beta$ ratio and energies used in previous X-ray phase imaging studies of similar samples \cite{Beltran:2010}. These values were obtained from the Center for X-Ray Optics (CXRO) website \cite{CXROWebsite}. A regularization value of $\varepsilon  = 1.5\times 10^{7}$ was used, which is several orders in magnitude less than $\left | \textbf{q} \right |^{2}$ even near the origin.

\begin{figure}[h!]
\centering\includegraphics[width=13cm]{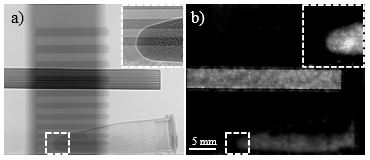}
\caption{Speckle-based transmission and diffusive dark-field retrieval results, for the sample shown in Fig.~\ref{Fig1}. (a) Image of the ratio $\textup{I}_{S}/\textup{I}_{R}$. (b) Diffusive dark-field image recovered using Eq.~(\ref{eq:EQ05}). The scale bar in (b) also applies to (a). The region outlined with a white dashed line is shown magnified in the top right of each panel. The dark-field image is shown with a linear grayscale from $\mathfrak{D} = 0$ to $\mathfrak{D} = 6.0 \times 10^{-11}$ ($\theta_{S}= 11 \upmu$rad), corresponding to $F= 1$ according to Eq.~(\ref{eq:PositionDependentScalarDiffusionCoefficient}).}\label{Fig3}
\end{figure}

Figure~\ref{Fig3}(a) shows a PBI-like image of the test sample, in which the region of the image behind the polystyrene micro-spheres yields a considerable level of residual speckle originating from the dark-field effects on the reference mask. We believe these artifacts arise where the term $\textup{I}_{S} / \textup{I}_{R}$ in Eq.~(\ref{eq:EQ02}) does not entirely remove the mask pattern, because the mask pattern has been substantially changed by the presence of the sample in these regions. The associated recovered diffusive dark-field image in Fig.~\ref{Fig3}(b) yields a strong signal in the regions where micro-spheres are present, but no signal from the surrounding solid plastic tube, indicating that this method is indeed highly sensitive to diffusive effects induced by SAXS/USAXS due to micro-structure. Note, also, that: (i) the stated value of $F=1$, as given in the caption to Fig.~\ref{Fig3}, is consistent with the fact that the Fresnel number, associated with propagation from the entrance surface to the exit surface of the sample, is much greater than unity \cite{Paganin2022Preprint}; (ii) the recovered position-dependent blur angles $\theta_{S}$ correspond to microstructure-induced blur at the detector plane over a maximum transverse distance on the order of twice the effective pixel size.  

The horizontally-placed wooden skewer also reveals a strong signal in the recovered dark-field. The inner structure of wood contains high levels of porosity and fibers, known to induce diffusive effects and therefore producing a strong diffusive dark-field signal. Interestingly, incorporating a wood skewer undoubtedly violates the premise of the single-material sample restriction assumed in deriving this method, yet we still obtain a usable signal that is valuable for qualitative assessment.       

\subsection{Diffusive dark-field imaging using a single periodic grid}\label{subsec2}

In this section we apply our diffusive dark-field retrieval method to the same sample and setup, but now the random mask is replaced with a grid containing a two-dimensional periodic structure. The raw images of (i) the mask and (ii) the mask and sample are shown in Figs.~\ref{Fig1}(a-b), respectively. Note that all parameters of the experiment other than the mask remain unchanged, hence the same input values for $\delta$ and $\beta$ were used as in the previous section.   

\begin{figure}[h!]
\centering\includegraphics[width=13cm]{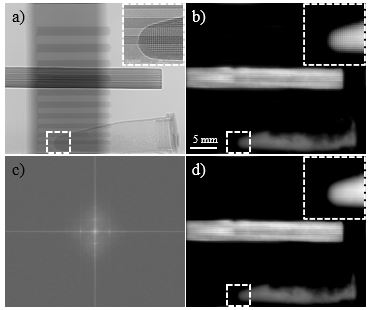}
\caption{Grid-based retrieved transmission and diffusive dark-field images, for the sample shown in Fig.~\ref{Fig1}. (a) The PBI-like image obtained from the ratio $\textup{I}_{S}/\textup{I}_{R}$. (b) The diffusive dark-field image recovered using Eq.~(\ref{eq:EQ05}). (c) The Fourier transform of (b), shown as the logarithm of the absolute value of the complex image, showing the grid artifacts as harmonic peaks. (d) The retrieved diffusive dark-field image, where a Fourier filter has removed the grid artifacts seen in (b). The scale bar in (b) also applies to (a) and (d). The dark-field images are shown with a linear grayscale from $\mathfrak{D} = 0$ to $\mathfrak{D} = 7.2 \times 10^{-10}$ ($\theta_{S}= 38$  $\upmu$rad), corresponding to $F= 1$ according to Eq.~(\ref{eq:PositionDependentScalarDiffusionCoefficient}). }\label{Fig4}
\end{figure}

Diffusive dark-field images retrieved from data collected with a grid-based mask are shown in Fig.~\ref{Fig4}. Immediately one can notice residual periodic artifacts from the grid in the transmission image (Fig.~\ref{Fig4}(a)) in the regions where micro-structure is found, in the same way residual speckles are seen in these regions when using the speckle mask. These artifacts propagate to some degree to the recovered dark-field image (inset, Fig.~\ref{Fig4}(b)). However because the differences are local, these high spatial frequencies are reduced by the Fourier filter used in the transmission retrieval (Eq.~(\ref{eq:EQ02})), and this does not prevent the successful application of the algorithm. As seen in the previous section, the retrieved diffusive dark-field image in Fig.~\ref{Fig4}(b) displays high sensitivity to areas where local blurring takes place, from both the polystyrene micro-spheres and the wooden fibers.     

To remove the periodic artifacts we utilized an image processing method which blocks the periodicity signatures in Fourier space, by locally setting them to zero. These periodicity signatures are clearly visualized as the bright spots in the magnitude of the Fourier transform of Fig.~\ref{Fig4}(b), shown here as Fig.~\ref{Fig4}(c). Once these harmonic peaks are blocked, using a computational form of the technique of spatial filtering (see e.g.~pp~642--647 of Ref.~\cite{HechtOpticsBook}), the image can be inverse Fourier transformed back to real space.  In this way, the artifacts can be almost completely suppressed, as shown in Fig.~\ref{Fig4}(d).  

\subsection{Testing the stability under varying noise levels}\label{subsec3}

\begin{figure}[h!]
\centering\includegraphics[width=13cm]{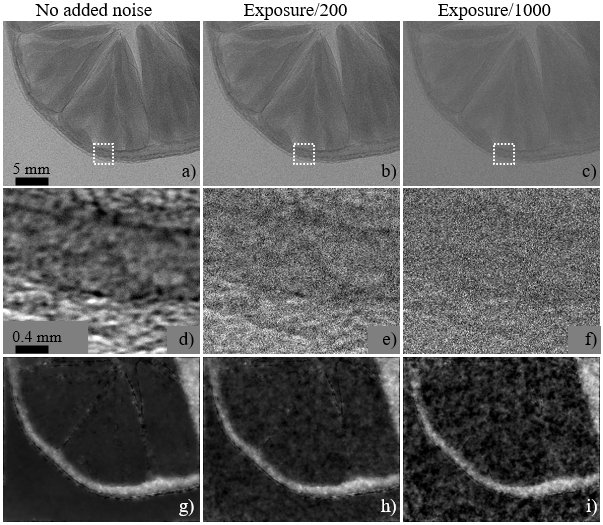}
\caption{Raw random mask-based images $\textup{I}_{S}$ of a lemon sample are shown with increasing levels of Poisson noise, where (b) has $7\%$ of the SNR of (a), and (c) $3\%$ of the SNR of (a). Magnified regions of (a), (b) and (c) are seen in (d), (e) and (f), respectively. Retrieved diffusive dark-field images for the different noise levels are shown in (g), (h) and (i). The scale bar in (a) applies to all images except for those in the middle row. 
 The dark-field images are shown with a linear grayscale from $\mathfrak{D} = 0$ to $\mathfrak{D} = 4.2\times 10^{-11}$ ($\theta_{S}= 9.1  \upmu$rad), corresponding to $F= 1$ according to Eq.~(\ref{eq:PositionDependentScalarDiffusionCoefficient}). }\label{Fig5}
\end{figure}

Despite the crude assumptions made as a trade-off in order to reach mathematical simplicity and computational efficiency, we note another benefit---in addition to computational speed---namely the robustness of the method with respect to noise. Accordingly, we tested the stability of the retrieval algorithm on random-mask-based data of a lemon slice (see Fig.~\ref{Fig5}(a)). The experimental parameters and setup were the same as in the previous two sections. Poisson noise was added to the raw image $\textup{I}_{S}$ prior to applying our retrieval approach. The increasing added noise in the raw images is shown in Figs.~\ref{Fig5}(a-c) but is considerably more obvious in Figs.~\ref{Fig5}(d-f), which are magnified regions, shown by the square boxed area in the first row of the figure. Signal-to-noise ratios ($\textup{SNR} = \sqrt{N}$, with $N$ defined below) of the input images in the top row were calculated to be 107 for (a), 7.6 for (b) and 3.4 for (c), indicating an effective lower average photon count per pixel with decreasing SNR as the added noise level is increased. $N=11\,572$ is the mean gray value of the original image in (a), which was varied as $N/200$ for (b) and $N/1000$ for (c). For the respective added noise levels, diffusive dark-field reconstructions are shown in Figs.~\ref{Fig5}(g-i). Evidently, the higher the level of added noise, the poorer the diffusive dark-field reconstruction, yet the SAXS/USAXS signal is still easily distinguished in all cases. In terms of practicality, this extreme noise stability can prove beneficial when it comes to radiation-dose considerations, especially when dealing with biological specimens. 

\section{Discussion and conclusion}\label{sec4}

We have derived a fast implicit diffusive-dark-field retrieval algorithm for data measured using a single position of a mask, which may be periodic or random. The main advantage of single-mask imaging is that only one exposure of the sample is required for the data acquisition process. This makes the technique highly suitable for high-speed dynamic imaging, compared to the other mask-based diffusive dark-field X-ray retrieval methods which require multiple sample exposures. Note also that because the mask is stationary, the retrieval algorithm should be robust to any motion blur due to a moving sample, since this would not blur the mask and hence would not be interpreted as dark-field. 

The primary advantage of our method is its robustness in retrieving SAXS/USAXS signal pertaining to sample unresolved micro-structure. Propagation-based phase contrast effects are incorporated within this approach, and hence do not appear as edge signal in the retrieved dark-field images. This means that the resulting images extract the diffusive type scattering signal. Another benefit of the approach is high stability under shot noise whilst only utilizing one exposure. This makes it potentially useful for low-dose imaging using laboratory-based sources that have considerably less photon flux than synchrotron sources. We believe the robustness and sensitivity will be maximized when using a high-visibility pattern, as reflected in Figs.~\ref{Fig3} and \ref{Fig4}, where the grid pattern enables a lower-noise diffusive dark-field retrieval than the lower-visibility speckle pattern does. The sensitivity of the set-up can be tuned by adjusting the sample-to-detector distance, such that the reference pattern is appreciably blurred, but not entirely blurred out. While we have found this optimization to be most easily performed experimentally, the propagation distance should not be so long that the linearization of the Fokker-Planck equation is no longer valid (for detail, see Eq.~(30) in Ref.~\cite{Paganin2022Preprint}).           
While there are reference-pattern remnants in areas of strong diffusive dark-field in the ``transmission'' images in panel a) of Figs.~\ref{Fig3} and \ref{Fig4}, this contrast is effectively the source of our dark-field contrast. If someone wished to retrieve a transmission image without these remnants, the retrieved dark-field image (e.g.~Fig.~\ref{Fig4}(d)) could be used to regionally blur the reference image, $\textup{I}_R$, before calculating the transmission $\textup{t}_0$ using Eq.~(\ref{eq:EQ02}). Future work could potentially introduce an iteration to the algorithm where the resulting refined transmission image $\textup{t}_0$ is used to improve the estimate of the flow term (Eq.~(\ref{eq:EQ04})) and hence potentially reduce the reference pattern remnants seen in the dark-field images.

In terms of drawbacks of the approach, the assumption that the object is composed of a single material clearly restricts the classes of specimen this method can be applied to. Despite this limitation, we saw that even though the samples used in this investigation were not entirely homogeneous, the reconstruction showed a high sensitivity to the wood interfaces placed within the main object, thus providing extremely useful visual information indicating that these produce a high degree of diffusive signal. This opens up the idea that this method could be extended to a wider class of samples (non-homogeneous) where one could focus in on a particular material interface by tuning the values of the refractive index corresponding to the material of interest and at the same time still only using one exposure. This would be analogous to what has been done in studies by Gureyev $et$ $al$.~\cite{Gureyev2materials:2002} in the context of quantitative material-specific X-ray phase retrieval using PBI \cite{Beltran:2010}. In the case that the sample thickness retrieval step (Eq.~(\ref{eq:EQ02})) over- or under-smooths some material interfaces within a multi-material sample, we expect any resulting artifacts in the dark-field image to simply highlight edges. Further to this, another particular class of samples that would be interesting to consider in future studies are those classified as pure phase objects ($\textup{t}_{0} = 1$). In this case estimation of the transmission would not be needed and the single-material assumption could be dropped, with sample thickness being retrieved from the phase shift by tracking speckles \cite{Paganin:2018}. Another possible avenue of research is to make use of the retrieved SAXS/USAXS signal as an initial guess (a seed) to more computationally-intensive iterative diffusive dark-field reconstruction approaches where no restrictions on the sample are made. Studying the influence of the nonzero distance, between the mask and the object, is another interesting avenue for future work.

It is also worth commenting on the possible influence of long-range tails in the scattering function, namely particularly slow transverse spatial decay associated with the position-dependent SAXS/USAXS distribution functions.  If such long-range tails are present, then the first term on the right side of Eq.~(\ref{eq:EQ01}) will need to be augmented with higher-order spatial derivatives.  Our position-dependent diffusion coefficient will then become the first member in an infinite hierarchy of diffusion tensors, with our Fokker--Planck equation thereby generalizing to an equation of the Kramers--Moyal form.  The associated higher-order diffusion coefficients are closely related to moments of the local SAXS/USAXS distribution.  For more information on these points, we refer to our earlier papers on the application of the Fokker--Planck equation to paraxial optics \cite{PaganinMorgan:2019,MorganPaganin:2019}, together with references contained therein.   

Finally, we believe this method could be applied to data measured from single-mask setups where the probing radiation is comprised of matter waves such as neutrons and electrons. We base this statement on the fact that our underpinning expression, namely Eq.~(\ref{eq:EQ01}), is in essence a statement of local energy conservation under the simultaneous presence of both coherent and diffusive flow, together with the presence of a sink term associated with a dimensionless sample-transmission function. Such local energy conservation, embodied in a suitable reverse-flow Fokker--Planck form of the continuity equation, holds for a very broad class of matter and radiation fields.

\section*{Acknowledgments}

The authors are grateful for help provided by the beamline scientists, Chris Hall, Daniel Hausermann, Anton Maksimenko and Matthew Cameron, at the Imaging and Medical beamline at the Australian Synchrotron, part of ANSTO, where the experiment was completed under proposal 18642. KM acknowledges support from the Australian Research Council (FT18010037) and MC acknowledges funding from the Australian Research Training Program (RTP). 

\section*{Disclosures}
The authors declare no conflicts of interest.

\section*{Data availability}
The raw and diffusive-dark-field retrieved images presented in this paper can be obtained from \url{https://github.com/Xray-grill/DiDaFi_MKCroughan}


\bibliographystyle{unsrt}  
\bibliography{Manuscript}

\begin{thebibliography}{10}

\bibitem{Bravin2013}
Alberto Bravin, Paola Coan, and Pekka Suortti.
\newblock X-ray phase-contrast imaging: from pre-clinical applications towards
  clinics.
\newblock {\em Physics in Medicine and Biology}, 58:R1, 2013.

\bibitem{FosterABI1980}
E.~F{\"o}rster, K.~Goetz, and P.~Zaumseil.
\newblock Double crystal diffractometry for the characterization of targets for
  laser fusion experiments.
\newblock {\em Kristall und Technik}, 15:937–945, 1980.

\bibitem{momose2003}
Atsushi Momose, Shinya Kawamoto, Ichiro Koyama, Yoshitaka Hamaishi, Kengo
  Takai, and Yoshio Suzuki.
\newblock Demonstration of x-ray {T}albot interferometry.
\newblock {\em Jpn J. Appl. Phys.}, 42(7B):L866, 2003.

\bibitem{WeitKamp2005}
Timm Weitkamp, Ana Diaz, Christian David, Franz Pfeiffer, Marco Stampanoni,
  Peter Cloetens, and Eric Ziegler.
\newblock X-ray phase imaging with a grating interferometer.
\newblock {\em Optics Express}, 16:6296--6304, 2005.

\bibitem{FranzPfeiffer2006LowBrilliancePC}
Franz Pfeiffer, Timm Weitkamp, Oliver Bunk, and Christian David.
\newblock Phase retrieval and differential phase-contrast imaging with
  low-brilliance x-ray sources.
\newblock {\em Nature Physics}, 2:258–261, 2006.

\bibitem{PFEIFERgrating:2006}
Franz Pfeiffer, Martin Bech, Oliver Bunk, Philipp Kraft, Eric~F. Eikenberry,
  Ch~Brönnimann, Christian Grünzweig, and Christian David.
\newblock Hard-x-ray dark-field imaging using a grating interferometer.
\newblock {\em Nature Materials}, 7:134--137, 2008.

\bibitem{OLIVOedge:2007}
Alessandro Olivo and Robert Speller.
\newblock A coded-aperture technique allowing x-ray phase contrast imaging with
  conventional sources.
\newblock {\em Applied Physics Letters}, 91:074106, 2007.

\bibitem{Endrizzi2014}
Marco Endrizzi, Paul~C. Diemoz, Thomas~P. Millard, J.~Louise Jones, Robert~D.
  Speller, Ian~K. Robinson, and Alessandro Olivo.
\newblock Hard x-ray dark-field imaging with incoherent sample illumination.
\newblock {\em Applied Physics Letters}, 104:024106, 2014.

\bibitem{Wen2010}
Harold~H. Wen, Eric~E. Bennett, Rael Kopace, Ashley~F. Stein, and Vinay Pai.
\newblock Single-shot x-ray differential phase-contrast and diffraction imaging
  using two-dimensional transmission gratings.
\newblock {\em Optics Letters}, 35:1932--1934, 2010.

\bibitem{KMorganGrid:2011}
Kaye~S. Morgan, David~M. Paganin, and Karen K.~W. Siu.
\newblock Quantitative single-exposure x-ray phase contrast imaging using a
  single attenuation grid.
\newblock {\em Optics Express}, 19(20):19781--19789, 2011.

\bibitem{BerujonSpeckle:2012}
Sébastien Bérujon, Eric Ziegler, Roberto Cerbino, and Luca Peverini.
\newblock Two-dimensional x-ray beam phase sensing.
\newblock {\em Physical Review Letters}, 108(15):158102, 2012.

\bibitem{KMorgan:2012}
Kaye~S. Morgan, David~M. Paganin, and Karen K.~W. Siu.
\newblock X-ray phase imaging with a paper analyzer.
\newblock {\em Applied Physics Letters}, 100:124102, 2012.

\bibitem{Cloetens1996}
Peter Cloetens, Raymond Barrett, Jos{\'e} Baruchel, Jean-Pierre Guigay, and
  Michel Schlenker.
\newblock Phase objects in synchrotron radiation hard x-ray imaging.
\newblock {\em Journal of Physics D: Applied Physics}, 29:133--146, 1996.

\bibitem{SnigerevPBI1995}
Anatoly Snigirev, Irina Snigireva, V.~Kohn, S.~Kuznetsov, and I.~Schelokov.
\newblock On the possibilities of x‐ray phase contrast microimaging by
  coherent high‐energy synchrotron radiation.
\newblock {\em Review of Scientific Instruments}, 66:5486, 1995.

\bibitem{SWILKINS1996}
S.~W. Wilkins, T.~E. Gureyev, D.~Gao, A.~Pogany, and A.~W. Stevenson.
\newblock Phase-contrast imaging using polychromatic hard x-rays.
\newblock {\em Nature}, 384:335–338, 1996.

\bibitem{Berujon2010Multi}
Sebastien Berujon, Hongchang Wang, and Kawal Sawhney.
\newblock X-ray multimodal imaging using a random-phase object.
\newblock {\em Physical Review A}, 86:063813, 2012.

\bibitem{Morgan2013Grid}
Kaye~S. Morgan, Peter Modregger, Sarah~C. Irvine, Simon Rutishauser, Vitaliy~A.
  Guzenko, Marco Stampanoni, and Christian David.
\newblock A sensitive x-ray phase contrast technique for rapid imaging using a
  single phase grid analyzer.
\newblock {\em Optics Letters}, 38:4605--4608, 2013.

\bibitem{Zanette:2014}
I.~Zanette, T.~Zhou, A.~Burvall, U.~Lundstr{\"o}m, D.H. Larsson, M.~Zdora,
  P.~Thibault, F.~Pfeiffer, and H.M. Hertz.
\newblock Speckle-based x-ray phase-contrast and dark-field imaging with a
  laboratory source.
\newblock {\em Physical Review Letters}, 112:253903, 2014.

\bibitem{Zdora:UMPA}
Marie~Christine Zdora, Pierre Thibault, Tunhe Zhou, Frieder~J. Koch, Jenny
  Romell, Simone Sala, Arndt Last, Christoph Rau, and Irene Zanette.
\newblock X-ray phase-contrast imaging and metrology through unified modulated
  pattern analysis.
\newblock {\em Physical Review Letters}, 118(3):203903, 2017.

\bibitem{TorbenDDF2010}
Torben~H. Jensen, Martin Bech, Oliver Bunk, Tilman Donath, Christian David,
  Robert Feidenhans'l, and Franz Pfeiffer.
\newblock Directional x-ray dark-field imaging.
\newblock {\em Physics in Medicine and Biology}, 55:3317--23, 2010.

\bibitem{Jensen2010b}
Torben~Haugaard Jensen, Martin Bech, Irene Zanette, Timm Weitkamp, Christian
  David, Hans Deyhle, Simon Rutishauser, Elena Reznikova, J\"urgen Mohr, Robert
  Feidenhans'l, and Franz Pfeiffer.
\newblock Directional x-ray dark-field imaging of strongly ordered systems.
\newblock {\em Phys. Rev. B}, 82:214103, Dec 2010.

\bibitem{TUNHE2018}
Tunhe Zhou, Hongchang Wang, and Kawal Sawhney.
\newblock Single-shot x-ray dark-field imaging with omnidirectional sensitivity
  using random-pattern wavefront modulator.
\newblock {\em Applied Physics Letters}, 113:091102, 2018.

\bibitem{FlorianSchaff2017}
Florian Schaff, Friedrich Prade, Yash Sharma, Martin Bech, and Franz Pfeiffer.
\newblock Non-iterative directional dark-field tomography.
\newblock {\em Scientific Reports}, 7:03307, 2017.

\bibitem{dreier2020}
Erik~S Dreier, Chantal Silvestre, Jan Kehres, Daniel Turecek, Mohamad Khalil,
  Jens~H Hemmingsen, Ole Hansen, Jan Jakubek, Robert Feidenhans'l, and Ulrik~L
  Olsen.
\newblock Single-shot, omni-directional x-ray scattering imaging with a
  laboratory source and single-photon localization.
\newblock {\em Optics Letters}, 45(4):1021--1024, 2020.

\bibitem{smith2022}
Ronan Smith, Fabio De~Marco, Ludovic Broche, Marie-Christine Zdora, Nicholas~W
  Phillips, Richard Boardman, and Pierre Thibault.
\newblock X-ray directional dark-field imaging using unified modulated pattern
  analysis.
\newblock {\em PLoS One}, 17(8):e0273315, 2022.

\bibitem{berujon2015}
Sebastien Berujon and Eric Ziegler.
\newblock Near-field speckle-scanning-based x-ray imaging.
\newblock {\em Physical Review A}, 92(1):013837, 2015.

\bibitem{YingAndMorgan:2022}
Ying~Y. How and Kaye~S. Morgan.
\newblock Quantifying the x-ray dark-field signal in single-grid imaging.
\newblock {\em Optics Express}, 30:10899--10918, 2022.

\bibitem{Paganin:2018}
David~M. Paganin, Hélène Labriet, Emmanuel Brun, and Sebastien Berujon.
\newblock Single-image geometric-flow x-ray speckle tracking.
\newblock {\em Physical Review A}, 98:053813, 2018.

\bibitem{Pavlov:2020}
Konstantin~M. Pavlov, Heyang. Li, David~M. Paganin, Sebastien Berujon, Helene
  Rouge-Labriet, and Emmanuel Brun.
\newblock Single-shot x-ray speckle-based imaging of a single-material object.
\newblock {\em Physical Review Applied}, 13(3):054023, 2020.

\bibitem{LaureneQuenot:2021}
Laur\`{e}ne Qu\'{e}not, Helene Roug\'{e}-Labriet, Sylvain Bohic, Sebastien
  Berujon, and Emmanuel Brun.
\newblock Implicit tracking approach for x-ray phase-contrast imaging with a
  random mask and a conventional system.
\newblock {\em Optica}, 8:111412--04, 2021.

\bibitem{Pavlov:2020B}
Konstantin~M Pavlov, David~M Paganin, Heyang~(Thomas) Li, Sebastien Berujon,
  Hélène Rougé-Labriet, and Emmanuel Brun.
\newblock X-ray multi-modal intrinsic-speckle-tracking.
\newblock {\em Journal of Optics}, 22(12):125604, nov 2020.

\bibitem{PavlovDdark:2021}
Konstantin~M. Pavlov, David~M. Paganin, Kaye~S. Morgan, Heyang~(Thomas) Li,
  Sebastien Berujon, Laurène Quénot, and Emmanuel Brun.
\newblock Directional dark-field implicit x-ray speckle tracking using an
  anisotropic-diffusion {F}okker-{P}lanck equation.
\newblock {\em Physical Review A}, 104(15):053505, 2021.

\bibitem{SAlloo:2022}
Samantha~J. Alloo, David~M. Paganin, Kaye~S. Morgan, Marcus~J. Kitchen,
  Andrew~W. Stevenson, Sheridan~C. Mayo, Heyang~T. Li, Ben~M. Kennedy, Anton
  Maksimenko, Joshua~C. Bowden, and Konstantin~M. Pavlov.
\newblock Dark-field tomography of an attenuating object using intrinsic x-ray
  speckle tracking.
\newblock {\em Journal of Medical Imaging}, 9(3):031502--14, 2022.

\bibitem{PaganinMorgan:2019}
David~M. Paganin and Kaye~S. Morgan.
\newblock X-ray {F}okker–{P}lanck equation for paraxial imaging.
\newblock {\em Scientific Reports}, 9(1):17537, 2019.

\bibitem{MorganPaganin:2019}
Kaye~S. Morgan and David~M. Paganin.
\newblock Applying the {F}okker–{P}lanck equation to grating-based x-ray
  phase and dark-field imaging.
\newblock {\em Scientific Reports}, 9(1):17465, 2019.

\bibitem{Teague1983}
Michael~Reed Teague.
\newblock Deterministic phase retrieval: a {G}reen’s function solution.
\newblock {\em Journal of the Optical Society of America}, 73:1434--1441, 1983.

\bibitem{Paganin2006}
D.~M. Paganin.
\newblock {\em Coherent X-{R}ay Optics}.
\newblock Oxford University Press, Oxford, 2006.

\bibitem{PaganinPelliccia2020}
D.~M. Paganin and D.~Pelliccia.
\newblock {X}-ray phase-contrast imaging: a broad overview of some
  fundamentals.
\newblock {\em Adv. Imaging Electron Phys.}, 218:63--158, 2021.

\bibitem{Paganin2022Preprint}
David~M. Paganin, Daniele Pelliccia, and Kaye~S. Morgan.
\newblock Paraxial diffusion-field retrieval, 2023.
\newblock arXiv:2301.09046.

\bibitem{Paganin:2002}
David Paganin, Sherry~C. Mayo, Timothy~E. Gureyev, P.~R. Miller, and Stephen~W.
  Wilkins.
\newblock Simultaneous phase and amplitude extraction from a single defocused
  image of a homogeneous object.
\newblock {\em Journal of Microscopy}, 206(3):33--40, 2002.

\bibitem{Paganin:2004}
D.~Paganin, T.~E. Gureyev, S.~C. Mayo, A.~W. Stevenson, {\relax{Ya}}.~I.
  Nesterets, and S.~W. Wilkins.
\newblock X-ray omni microscopy.
\newblock {\em Journal of Microscopy}, 214(3):315--327, 2004.

\bibitem{Gureyev2materials:2002}
Timothy~E. Gureyev, Andrew~W. Stevenson, David~M. Paganin, Timm Weitkamp,
  Anatoly Snigirev, Irina Snigireva, and Stephen~W. Wilkins.
\newblock Quantitative analysis of two-component samples using in-line hard
  x-ray images.
\newblock {\em Journal of Synchrotron Radiation}, 9:148--153, 2002.

\bibitem{Beltran:2010}
Mario~A. Beltran, David~M. Paganin, Kentaro Uesugi, and Marcus~J. Kitchen.
\newblock 2{D} and 3{D} x-ray phase retrieval of multi-material objects using a
  single defocus distance.
\newblock {\em Optics Express}, 18:6423--6436, 2010.

\bibitem{irvine2014}
S~Irvine, R~Mokso, P~Modregger, Z~Wang, F~Marone, and M~Stampanoni.
\newblock Simple merging technique for improving resolution in qualitative
  single image phase contrast tomography.
\newblock {\em Optics Express}, 22(22):27257--27269, 2014.

\bibitem{CXROWebsite}
{X-Ray Interactions With Matter}.
\newblock \url{https://henke.lbl.gov/optical_constants/}.
\newblock Hosted at website of Center for X-Ray Optics, Lawrence Berkeley
  National Laboratory (LBNL), Materials Sciences Division.

\bibitem{HechtOpticsBook}
E.~Hecht.
\newblock {\em Optics}.
\newblock Pearson, Boston, 5 edition, 2017.

\end{thebibliography}

\end{document}